\documentclass[reprint,
 amsmath,amssymb,
 aps,
]{revtex4-2}
\usepackage{comment}

\usepackage{hyperref}
\usepackage{graphicx}% Include figure files
\usepackage{dcolumn}% Align table columns on decimal point
\usepackage{bm}% bold math
\usepackage[utf8]{inputenc}
\usepackage[T1]{fontenc}
\usepackage{amsmath,amssymb,amsfonts}
\usepackage{physics}           % for \ket{}, \bra{}, \braket{}, etc.
\usepackage{graphicx}          % for figures
\usepackage{bm}                % bold math symbols
\usepackage{hyperref}          % for hyperlinks
\usepackage{xcolor} 
\usepackage{graphicx}
\usepackage{float}% optional: for colored text or links
\usepackage{microtype}         % improves spacing
\usepackage{mathtools}         % enhanced math (optional but useful)
\usepackage{comment}  
\usepackage{subfigure}
\usepackage{booktabs} 
\begin{document}

\preprint{APS/123-QED}

\title{ Information conservation relations for weak measurement and its reversal }% Force line breaks with \\

\author{Yusef Maleki}
 \email{maleki@tamu.edu}

 \author{Luis D. Zambrano Palma}
  \email{ldzambra@tamu.edu}

 \author{M. Suhail Zubairy}
 \email{zubairy@tamu.edu}

\affiliation{Institute for Quantum Science and Engineering, Texas A\&M University, College Station, Texas 77843, USA}

\date{\today}

\begin{abstract}

We investigate the information  distribution among different entities in the weak measurements protocol. Focusing on multilevel, decaying systems under continuous (no-click) monitoring, we derive exact, conservation-type information relations that hold for each outcome in the record. Analogous relations hold when an explicit reversal is applied, with the reversal success probability entering the relation. 
 We extend the framework to finite-count outcomes (arbitrary photon numbers) obtaining quantitative trade-offs that link information change in the weak measurement process to the entities to which the information is distributed.  These results provide a unified, outcome-resolved account of information flow in monitored open quantum dynamics and provide insight into a deeper understanding of open-system dynamics and its control.

\end{abstract}

\maketitle

\section{Introduction}

Measurement is central to all scientific methodologies, enabling hypothesis testing and theory validation, and knowledge growth.
In quantum mechanics as well, measurement enables us to gain information about a system, but it is intrinsically more subtle as gaining information perturbs the system state. This information–disturbance trade-off is at the heart of the theory and is crucial in applications of quantum theory in quantum information science, from secure communication to quantum sensing \cite{bennett2014quantum,nielsen2010quantum,maleki2021quantum}. In a projective (von Neumann) measurement \cite{vonNeumann1955}, the post-measurement state is an eigenstate of the measured observable, leading to the maximal knowledge gain at the cost of an irreversible loss of coherence. By contrast, weak measurements extract only partial information and induce proportionally smaller disturbance; by avoiding full collapse, they can still preserve coherence and, in some settings, allow (partial) recovery of the initial state \cite{PhysRevLett.97.166805,PhysRevLett.68.3424,kim2012protecting,korotkov2010decoherence,xiao2013protecting,PhysRevA.47.R4577}. In a broader setting, within the POVM framework, a more general class of measurements beyond the von Neumann measurement can be performed where the measurement strength can be tuned \cite{PhysRevB.60.5737,kim2009reversing}. In the weak measurement setting, even null outcomes, with the absence of a registered click in the detector, carry information \cite{MasashiBan_2001,terashima2011information,PhysRevA.80.033838}.

An important observation about weak measurements is that such a measurement can also be caused by system-environment interactions, leading to an unwanted loss of quantum coherence \cite{zurek2003decoherence,schlosshauer2004decoherence,maleki2018generating}. To be more specific, uncontrolled coupling to the environment damages superposition and entanglement, eliminating the non-classical features of quantum systems.
Considering the deleterious effect of decoherence as an important obstacle for scalable quantum computing and for performance of quantum technologies,  a significant attention was paid for characterizing, quantifying, and mitigating coherence loss across diverse platforms \cite{terhal2015quantum,cai2023quantum}. Within the resource-theoretic perspective \cite{baumgratz2014quantifying}, multiple measures of coherence have been introduced and analyzed, providing operational benchmarks for experiments \cite{knill1998resilient,viola1998dynamical}. In parallel, techniques such as quantum error correction \cite{terhal2015quantum}, reservoir engineering \cite{maleki2020maximal},  dynamical decoupling, and decoherence-free subspaces \cite{lidar2014review} aim to preserve and manipulate coherence in realistic settings. Progress on these fronts  is essential for turning proof-of-principle demonstrations into reliable quantum technologies.

Numerous studies have focused on understanding how information gain, disturbance, and the chance of reversing a weak measurement are quantitatively linked. No-click monitoring of decaying systems yields conditional evolution with mutual information that increases over time, while the success probability of recovery decreases \cite{Sun2009}. Relations connecting fidelity, mutual information,  reversibility, and information gain have been considered for general quantum measurements \cite{PhysRevA.92.022114,PhysRevA.73.042307,Terashima2016,Cheong2012}. Complementary analyses use entropic, fidelity-based, and coherence-based tools to obtain relations across measurement strengths \cite{PhysRevA.92.022114}. Further developments clarify the role of null outcomes, specify conditions for reversibility and recovery, and introduce relations for information flow in open dynamics \cite{Konishi2024,Nielsen1997,Liao2013,Roncaglia2019,Fullwood2025,Lee2019}. However, despite substantial progress, a complete quantitative account of information flow under weak measurements, especially in multilevel decaying systems, is not well understood.

In this paper, we address this by deriving conservation-type information relations that hold for each outcome in the measurement record. In particular, we demonstrate how the change in the system’s information content is distributed as contributions from null-result (no-click) information and the decay channel's informatics content. We then extend the framework to weak-measurement reversal, explicitly incorporating the information carried by the reversal success probability and show how a conservation-type information relation holds. The analysis is further generalized to finite-count outcomes, where the detector registers an arbitrary number of photons, resulting in a generalized conservation relation in the system. Together, these results provide a unified account of information flow in monitored open quantum dynamics and provide insight into a deeper understanding of open-system dynamics and its control.

\section{Informational Conservation Analysis}

\subsection{Single Qubit Setting }

To illustrate our information-theoretic analysis, we begin with the simplest case: a single qubit. For concreteness, we model it as a superposition of the vacuum and one-photon Fock states in a cavity, and we characterize the resulting information flow. Equivalently, one may view the qubit as a two-level atom coupled to a cavity mode and track its dynamics under the weak measurement. 
The state of a qubit inside of a cavity is expressed as
\begin{equation}
\ket{\psi} = c_0 \ket{0} + c_1 \ket{1} .
\end{equation}
For the weak measurement setting, a detector is placed outside the cavity to monitor photons leaking out of the cavity. For the case of a superposition of Fock states inside a cavity, the corresponding measurement operator, referred to as the photon-detection measurement operator \cite{ueda1989probability}, is expressed as
\begin{equation}
  M_k = \sqrt{\frac{(1-e^{-2\gamma t})^k}{k!}}\, e^{-\gamma t \hat{n}} \hat{a}^{k}. \label{wm}
\end{equation}
Here, \(\hat{a}\) is the annihilation operator and \(\hat{n}\) the number operator. For the qubit case, a null detection occurs when the detector does not register a click within a given time interval. The corresponding null-result operator \((k=0)\) is then
\begin{equation}
    M_0 = e^{-\gamma t \hat{n}} \label{wmo}.
\end{equation}
Therefore, after a weak measurement with null result, the field state evolves to
\begin{equation}
\ket{\psi'} = \frac{M_0 \ket{\psi}}{\sqrt{\bra{\psi} M_0^{\dagger} M_0 \ket{\psi}}}. \label{mse}
\end{equation}
In the particular case of a qubit, the state takes the form
\begin{equation}
\ket{\psi'} = \frac{ c_0 \ket{0} + c_1 e^{-\gamma t} \ket{1} }
{\sqrt{|c_0|^2 + |c_1|^2 e^{-2\gamma t}}}.
\end{equation}
where \(\gamma\) is the decay rate and \(t\) the evolution time. Note that if one considers a two-level atom inside a cavity, where $\ket{0}$ and $\ket{1}$ correspond to the logical encoding of the ground and excited states, the resultant state of the system after the weak measurement still takes the same form as above.

Therefore, the probability of finding the system in the ground state \(\ket{0}\) is $p(x_0) = |c_0|^2.$ Also,
the probability that the detector registers a null result (does not click) during a time interval \(t\) is
\begin{equation}\label{py0}
p(y_0) = \bra{\psi} M_0^{\dagger} M_0 \ket{\psi} 
       = |c_0|^2 + |c_1|^2 e^{-2\gamma t}.
\end{equation}

The probability of attaining state \(\ket{0}\) given that the detector registered no click (dark detection) is thus 

\begin{equation}
p(x_0|y_0) =\frac{|c_0|^2}{|c_0|^2 + |c_1|^2 e^{-2\gamma t}},
\end{equation}

Similarly, for the detection of \(\ket{1}\), we have $p(x_1)=|c_1|^2$ and the posterior is
\begin{equation}
p(x_1 | y_0)=\frac{|c_1|^2 e^{-2\gamma t}}{|c_0|^2+|c_1|^2 e^{-2\gamma t}}.
\end{equation}

Now, for the event \(a_n\) with the outcome probability $p(a_n)$, the information content associated with the event is $I(a_n) = -\log_2 p(a_n)$ 
\cite{Stenholm2005} . This  quantifies the information gained upon the observation of the event \(a_n\).

Therefore, considering the single-qubit case, the quantity $I(x_0)=-\log_2 p(x_0)$ is the information content of obtaining outcome $\ket{0}$ from the initial preparation. Likewise, $I(x_0\mid y_0)=-\log_2 p(x_0\mid y_0)$ is the information content of obtaining $\ket{0}$ conditioned on the weak-measurement outcome $y_0$ (dark detection).

We define the pointwise information gain as the difference between the prior information and the conditional information for the null outcome as
\begin{align}
    \Delta I(x_0|y_0) &= I(x_0) - I(x_0|y_0),
    \\
    \Delta I(x_1|y_0) &= I(x_1) - I(x_1|y_0).
\end{align}
Thus, the information variations are respectively given by 
\begin{align}\label{DIx0y0}
  \Delta I(x_0 | y_0)
= \log_2 \frac{1}{|c_0|^2+|c_1|^2 e^{-2\gamma t}},
\\
\Delta I(x_1| y_0)
= \log_2 \frac{e^{-2\gamma t}}{|c_0|^2+|c_1|^2 e^{-2\gamma t}}.
\end{align}

It is evident that, for a null result, $\Delta I(x_0 | y_0)\ge 0$ and $\Delta I(x_1 | y_0)\le 0$.
This, in fact, has an operational meaning: The positivity  of $\Delta I(x_0 | y_0)$ indicates that the null-result outcome increases the plausibility of the state to be in $x_0$ relative to its prior. In other words, it provides evidence for $\ket{0}$. Conversely, the negative value indicates that the null-result outcome  lowers the plausibility of $\ket{1}$, providing evidence against it. The magnitude $|\Delta I(x_n | y_0)|$ quantifies the strength of this evidence in bits.

In the small-time limit where $e^{-2\gamma t}\approx 1-2\gamma t$ is valid, we attain
\begin{equation}
\Delta I(x_0 | y_0)\approx \frac{2\gamma t\,|c_1|^2}{\ln 2}.
\end{equation}
Therefore, the evidence in favor of $x_0$ improves linearly with $t$ and with the initial excited-state probability $|c_1|^2$. This is expected, due to the fact that a null detection is less likely when the qubit most likely is in the excited state $\ket{1}$, since it tends to emit with a higher probability. So observing the null result provides stronger evidence (contains more information) for the ground state $\ket{0}$ when the initial excited-state probability $|c_1|^2$ is larger. 
Thus, the instantaneous information rate at $t=0$ is
\begin{equation}
\frac{d}{dt}\Delta I(x_0 | y_0)\Big|_{t=0}
= \frac{2\gamma\,|c_1|^2}{\ln 2}\ \text{bits/s}.
\end{equation}
When $|c_1|^2 \to 0$ the null detection does not provide information about $\ket{0}$, whereas $|c_1|^2 \to 1$ the information rate approaches its maximum $2\gamma/\ln 2$.

Similarly, for the small time limits with $e^{-2\gamma t}\approx 1-2\gamma t$, we have
\begin{equation}
\Delta I(x_1 | y_0)\approx \frac{-2\gamma t\,|c_0|^2}{\ln 2}.
\end{equation}

Now, we consider the relations between different information contents. First, 
 comparing Eq. \eqref{py0} and \eqref{DIx0y0}, we arrive at a relation linking the information gain from a null result to the information content of the no-click event as
\begin{equation}
I(y_0) = \Delta I(x_0|y_0) \label{eq12},
\end{equation}
Also, following a similar description and identifying \(p(\text{decay}) = e^{-2\gamma t}\), we arrive at
\begin{equation}\label{eq13sq}
    \Delta I(x_1|y_0) = I(y_0) - I(\text{decay}),
\end{equation}
where, $I(\text{decay})=-\log p(\text{decay})$.

We note that, in contrast to the ground-state case, the information content for the excited state involves the decay process. Hence, the information gain consists of two distinct contributions: one associated with the null result and the other with the decay dynamics.

\subsection{Multi-Level System Setting}

We now extend the analysis developed above to an \(n\)-level system. In this setting, a general quantum state in the Fock basis can be written as
\begin{equation}
\ket{\psi} = \sum_{n=0}^{N} c_n \ket{n},
\label{multi-level}
\end{equation}
subject to the same null-result measurement operator defined in Eq. \eqref{wmo}.
After a weak measurement, the general initial state evolves according to Eq. \eqref{mse} as
\begin{equation}
\ket{\psi'} = \frac{1}{\mathcal{N}} \sum_{n=0}^{N} c_n e^{-\gamma t n} \ket{n},
\end{equation}
where \(\gamma\) is the decay rate, \(t\) is the evolution time, and \(\mathcal{N}\) is the normalization factor associated with a null detection. The corresponding  null-detection probability at time \(t\) for a multi-level system is
\begin{equation}
p(y_0) = \bra{\psi} M_0^{\dagger} M_0 \ket{\psi} 
       = \sum_{n=0}^{N} |c_n|^2 e^{-2\gamma t n}.
\end{equation}

Analogous to the qubit case, the prior probability of the system being in the state \(\ket{n}\) is $p(x_n) = |c_n|^2.$ 
The probability of attaining state \(\ket{n}\) given that the detector registered no click (dark detection) is thus 

\begin{equation}
    p(x_n|y_0) = \frac{e^{-2\gamma t n} |c_n|^2}{\sum_{m=0}^{N} |c_m|^2 e^{-2\gamma t m}}.
\end{equation}

 For the null-result, the information gain can be defined as
\begin{equation}
\Delta I(x_n|y_0) = I(x_n) - I(x_n|y_0). 
\end{equation}
This is the pointwise information gain (in bits) about the system being in $x_n$ provided by observing the null outcome $y_0$.

With these elements, we find the information gain formula as
\begin{equation}
    \Delta I(x_n |y_0) = I(y_0) - n\,I(\text{decay}). 
    \label{eq19}
\end{equation}
 In particular, for \(n=0\) and \(n=1\), this equation reduces to Eq. \eqref{eq12} and Eq. \eqref{eq13sq}, respectively. Therefore, these equations are valid independently of $N$ in the multi-level system in Eq. \eqref{multi-level}.

Moreover, Eq. \eqref{eq19} can be equivalently written as
\begin{equation}
    I(y_0) = \Delta I(x_n | y_0) + n\, I(\text{decay}) ,
    \label{conservation1}
\end{equation}
Thus, equation \eqref{conservation1} expresses a conservation–type balance for the information carried by the null outcome $y_0$. The information $I(y_0)$ is decomposed into a pointwise information flow $\Delta I(x_n \mid y_0)$, which may be positive or negative, and a strictly nonnegative loss term $n\,I(\text{decay})$.  
In this description, the left–hand side is independent of $n$, while the right–hand side depends on $n$ explicitly. Therefore, the combination $\Delta I(x_n \mid y_0) + n\, I(\text{decay})$ is independent of $n$ (i.e., conserved).

\subsection{Average Information Gain in Multi-Level Systems}

We now extend the analysis by evaluating the average information gained from a null result, averaged over the distribution of conditional probabilities \(p(x_n|y_0)\). We define the average information gain as
\begin{equation}
\langle \Delta I(x_n|y_0) \rangle = \sum_{n} p(x_n|y_0)\, \Delta I(x_n|y_0). \label{eq13}
\end{equation}
It is illuminating to express $\Delta I(x_n | y_0)$ explicitly as
\[
\Delta I(x_n | y_0)
= I(x_n) - I(x_n | y_0)
= \log_2 \frac{p(x_n | y_0)}{p(x_n)} .
\]
Averaging over the posterior probability \(p(x_n|y_0)\) yields 
\[
\langle \Delta I(x_n|y_0) \rangle
= \sum_{n} p(x_n | y_0)\,\log_2 \frac{p(x_n | y_0)}{p(x_n)}.
\]
This is just the relative entropy between the initial and the final distribution, and hence, we have 
\begin{equation}
\langle \Delta I(x_n|y_0) \rangle=D\big(p(x_n|y_0) \mid \mid p(x_n)\big).
\end{equation}
From the positivity of the relative entropy, we can immediately conclude that $\langle \Delta I(x_n|y_0) \rangle \ge 0.$

By substituting this into Eq. \eqref{eq19}, and defining the averaged photon number after a null detection as $\langle n \rangle = \sum_{n} n \,p(x_n|y_0)=(-1/2\gamma)\partial_t\ln p(y_0)$, we find the conservation-type relation as

\begin{equation}
I(y_0)=D\big(p(x_n|y_0)\mid \mid p(x_n)\big)+ \langle n \rangle\, I(\text{decay}).
\label{eq14}
\end{equation}

Thus, we arrive at a compact relation, showing that the information associated with a null result of the weak measurement is conserved such that it is distributed as the relative entropy of the prior and posterior states and the  decay information.  

\begin{figure}[H]
  \centering
  \includegraphics[width=\linewidth]{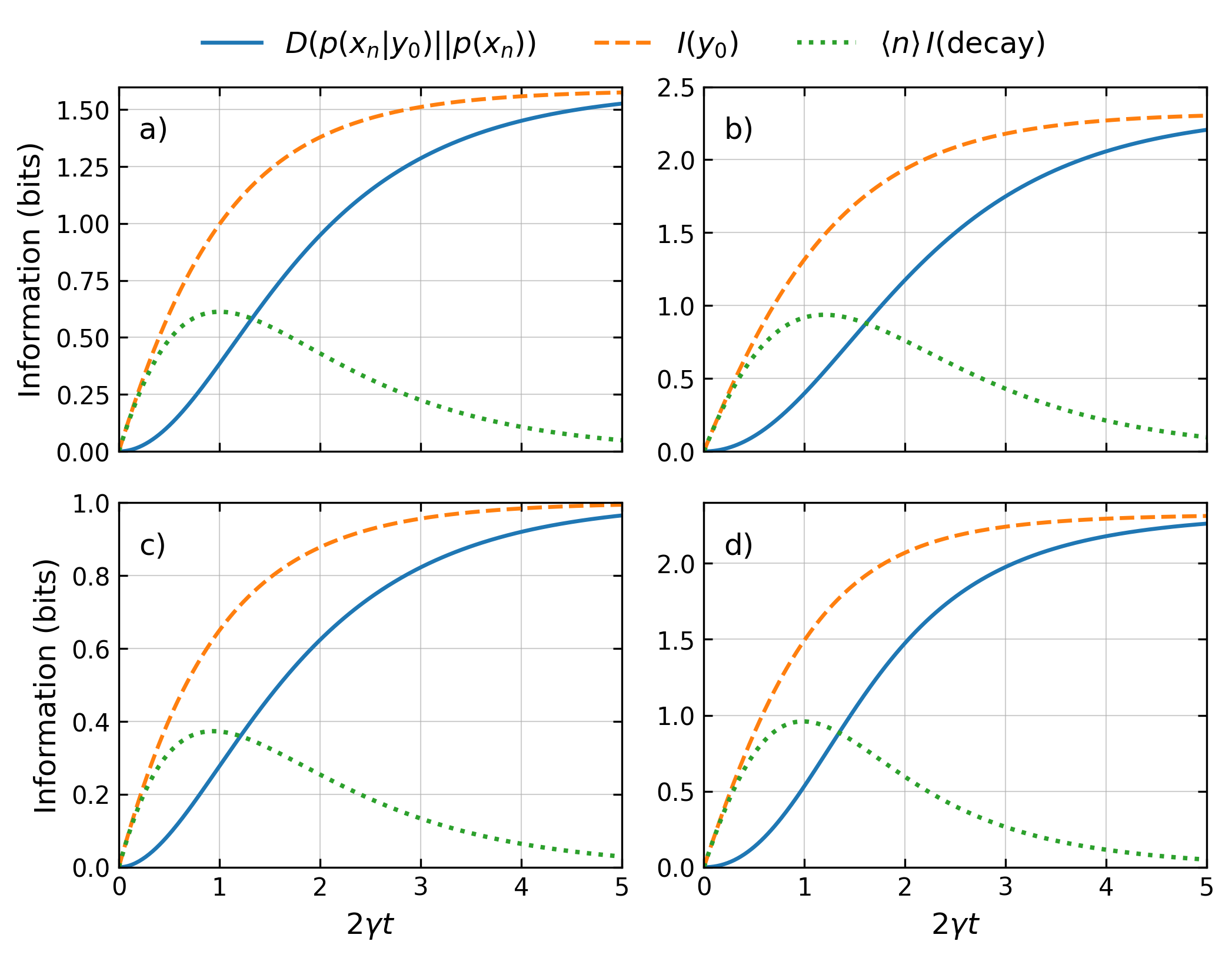}
\caption{
    Information-theoretic quantities as functions of $2\gamma t$ for four different prior distributions for the state \(\ket{\psi}=c_0\ket{0}+c_1\ket{1}+c_2\ket{2}\). 
    Each panel corresponds to different priors: a) $p(x_n) = [1/3,\,1/3,\,1/3]$, b) $p(x_n) = [0.2,\,0.4,\,0.4]$, c) $p(x_n) = [0.5,\,0.3,\,0.2]$, and d) $p(x_n) = [0.2,\,0.2,\,0.6]$.
  }
  \label{fig:eq14_terms}
\end{figure}

Fig. \ref{fig:eq14_terms} shows the averaged conservation-type expression for a three-level system (qutrit), considering different sets of initial coefficients for each state.  

As illustrated in Fig. \ref{fig:eq14_terms}, the information content of the null result, \(I(y_0)\), increases monotonically until it saturates at a limiting value. This limit corresponds to the information content of the initial state, i.e., \(-\log_2 p(x_0) \), where \(p(x_0) \) is the prior probability of the ground state.  The  contribution associated with $\langle n \rangle I(\text{decay})$ grows monotonically up to a maximum which occurs at $t=(1/2\gamma)[\langle n \rangle/\mathrm{Var}\langle n \rangle]$, where $\mathrm{Var}\langle n \rangle$ is the variance of $\langle n \rangle$. Beyond this point, it decreases monotonically and eventually vanishes, where no further decay information can be extracted.

Equation \eqref{eq14} immediately yields the bounds  
\begin{equation}
0 \leq D\big(p(x_n|y_0)\mid \mid p(x_n)\big) \leq I(y_0),
\end{equation}
with the upper limit reached for \(t = 0\).  

Physically, this means that the information gain from the dark count can never exceed the information content of relative entropy, quantifying the distinguishability between the initial and the final distributions. In other words, no-click events contain information and this information cannot exceed the distinguishability of the prior and posterior states quantified by the relative entropy.

\subsection{Reversal Probability and Information Balance}

 It is now natural to ask what role a possible reversal plays in a weak measurement protocol and if there is any conservation-type information relation which takes into account the information content of the reversal protocol.
 To this end, we follow the approach of Sun \emph{et al.} \cite{PhysRevA.80.033838} to characterize the reversibility by defining the reversal success probability as 
\begin{equation}
p (\text{rev}) =\frac{[p(\text{decay})]^N}{p(y_0)}.
\end{equation}
This expression gives the probability of restoring the system to its initial state after a null result in a weak measurement. This relation can be recast into an information-theoretic form as
\begin{equation}
    I(y_0) = N\,I(\text{decay}) - I(\text{rev}).
    \label{dsn}
\end{equation}
 This expression can be regarded as a conservation-type information relation for the probabilistic reversal process. Now, combining this relation with Eq. \eqref{eq19}, we have
\begin{equation}
I(y_0) = \frac{N\,\Delta I(x_n|y_0) + n\,I(\text{rev})}{N-n}.
 \label{dsn1}
\end{equation}

Although this form appears to depend explicitly on \(n\) and \(N\), its value is in fact constant for any level \(n\). This is due to the fact that \(I(y_0)\) is independent of the state index  \(n\). This provides a conservation-type expression: for each level \(n\), the combination of the information gain \(\Delta I(x_n|y_0)\) and the  reversal cost \(I(\text{rev})\) reconstructs the information associated with the null result. 
It therefore provides a different perspective on how the information content associated with null results can be distributed between gain and reversal. 

Since from Eq. \eqref{conservation1}, one has $I(y_0) = \Delta I(x_0|y_0) $, we can express Eq. \eqref{dsn} as
\begin{equation}
N\,I(\text{decay}) = I(\text{rev}) + \Delta I(x_0|y_0). 
\label{eq23}
\end{equation}
When we apply the same substitution into Eq. \eqref{dsn1}, we find
\begin{equation}
\Delta I(x_n|y_0) = \frac{N-n}{N}\,\Delta I(x_0|y_0) - \frac{n}{N}\,I(\text{rev}). 
\label{eq2}
\end{equation}

We can also rewrite Eq. \eqref{eq2}, for \(t>0\), in the form
\begin{equation}
\frac{n}{N} = \frac{\Delta I(x_0|y_0) - \Delta I(x_n|y_0)}{I(\text{rev}) + \Delta I(x_0|y_0)}. 
\label{suc2}
\end{equation}
A key aspect is that it connects the ratio of information quantities to the ratio of the excitation level \(n\) relative to the maximum level \(N\). 

As a next step, we evaluate Eq. \eqref{dsn1} at \(n=1\) and combine it with Eq. \eqref{eq23}, which gives
\begin{equation}
I(\text{rev}) = (N-1)\,\Delta I(x_0|y_0) - N\,\Delta I(x_1|y_0). 
\label{suc1}
\end{equation}
This relation shows how the information relevant to the reversal process after a null result is distributed. Notably, only the information gain from the ground state and from the first excited state, together with the system size \(N\), appears in the expression. Thus, even for systems with many higher levels, the reversal cost can be determined entirely from the contributions of the lowest two levels and the dimensionality of the state.

From Eqs. \eqref{suc1} and \eqref{eq2} we find that
\begin{equation}
\Delta I(x_n|y_0) = n\,\Delta I(x_1|y_0) - (n-1)\,\Delta I(x_0|y_0).
\end{equation}
Thus, the information gain at level \(n\) can be written as a linear combination of the contributions from the ground state and the first excited state. In other words, the informational structure of higher levels is fully determined by the two lowest levels.

Now, we average the information gain in Eq. \eqref{eq2}, similar to our earlier discussion. Hence, using the definitions in Eq. \eqref{eq13}, we obtain
\begin{equation}
\langle \Delta I(x_n|y_0) \rangle = \frac{N-\langle n \rangle}{N}\,\Delta I(x_0|y_0) - \frac{\langle n \rangle}{N}\,I(\text{rev}). 
\label{av3}
\end{equation}
or equivalently,
\begin{equation}
 I(y_0)=\frac{N}{N-\langle n \rangle} D\big(p(x_n|y_0)\mid \mid p(x_n)\big)+ \frac{\langle n \rangle}{N-\langle n \rangle}\,I(\text{rev}). 
\label{av3}
\end{equation}

Thus, a convex combination of the relative entropy and the reversal information \(I(\text{rev})\), gives the dark-count information. Note that one can alternatively rearrange this equation to express $I(\text{rev})$ in terms of the relative entropy and $ I(y_0)$.

\subsection{Information Gain in k-Click Events}

Next, we investigate how information is distributed for the setting where \(k\)-click events are allowed. Do the main relations in the null result case still hold their conservation-type structure? 
Therefore, we generalize the discussion to the case of an arbitrary outcome \(k\), corresponding to \(k\)-click events in the detector, following \cite{PhysRevA.80.033838}. In this case,  the probability of obtaining \(k\)-clicks during the weak measurement is computed using the measurement operator in Eq. \eqref{wm} as $p(y_k)= \bra{\psi} M_k^{\dagger} M_k \ket{\psi}$, resulting in
\begin{equation}
\begin{split}
p(y_k) = \sum_{n=0}^{N} \binom{n}{k}\,(1-e^{-2\gamma t})^k\,
          e^{-2\gamma (n-k)t}\,|c_n|^2 .
\end{split} \label{eq39a}
\end{equation}
Also, the conditional probability of observing \(k\)-clicks given that the system was initially in \(\ket{n}\) is $p(y_k|x_n)= \bra{n} M_k^{\dagger} M_k \ket{n}$. Hence
\begin{equation}
\begin{split}
p(y_k|x_n) = \binom{n}{k}\,(1-e^{-2\gamma t})^k\,e^{-2\gamma (n-k)t}.
\end{split}
\label{eq33}
\end{equation}

From Bayes’ theorem $p(x_n|y_k) = {p(x_n)\,p(y_k|x_n)}/{p(y_k)}$
 we obtain
\begin{equation}
p(x_n|y_k) = 
\frac{\binom{n}{k}(1-e^{-2\gamma t})^k\, e^{-2\gamma (n-k)t}\,|c_n|^2}
{\sum_{m=0}^{N} \binom{m}{k}(1-e^{-2\gamma t})^k\, e^{-2\gamma (m-k)t}\,|c_m|^2}.
\end{equation}
From this expression, the generalized unaveraged information gain for \(k\)-click events takes the form
\begin{equation}
\begin{split}
 I(y_k) ={}& \Delta I(x_n|y_k) \\
 &+ ( n  - k)\,I(\text{decay}) \\
 &+ k\,I(\text{no decay})
 -  I(W),  
\end{split}
\label{decaylaw}
\end{equation}
where $p(\text{no decay}) = 1 - e^{-2\gamma t}$, $W = \binom{n}{k}$ and $I(W)=\log_2\,W$.
It is clear that Eq. \eqref{decaylaw} generalizes the previous result of Eq. \eqref{eq19}, recovering the results for $k=0$. In particular, it contains the information associated with decay, its counterpart from the no-decay event, the contribution from detecting \(k\)-clicks, and an additional term that can be interpreted as the information cost of indistinguishability in the detection process. This conservation-type information relation therefore clarifies how the total information is distributed among different information elements.

To complete the generalized analysis, as in Eq. \eqref{eq13}, we average Eq. \eqref{decaylaw} over the posterior distribution \(p(x_n|y_k)\), which yields
\begin{equation}
\begin{split}
 I(y_k) ={}& D\!\big(p(x_n|y_k)\,\Vert\,p(x_n)\big) \\
 &+ (\langle n \rangle - k)\,I(\text{decay}) \\
 &+ k\,I(\text{no decay})
 - \langle I(W) \rangle .
 \label{k-click}
\end{split}
\end{equation}
where we define 
\begin{subequations}\label{eq:averages}
\begin{align}
\langle I(W)\rangle &= \sum_n I(W)\,p(x_n|y_k),  \\
\langle n \rangle &= \sum_{n} n \,p(x_n|y_k). 
\end{align}
\end{subequations}

\begin{figure}[H]
  \centering
  \includegraphics[width=\linewidth]{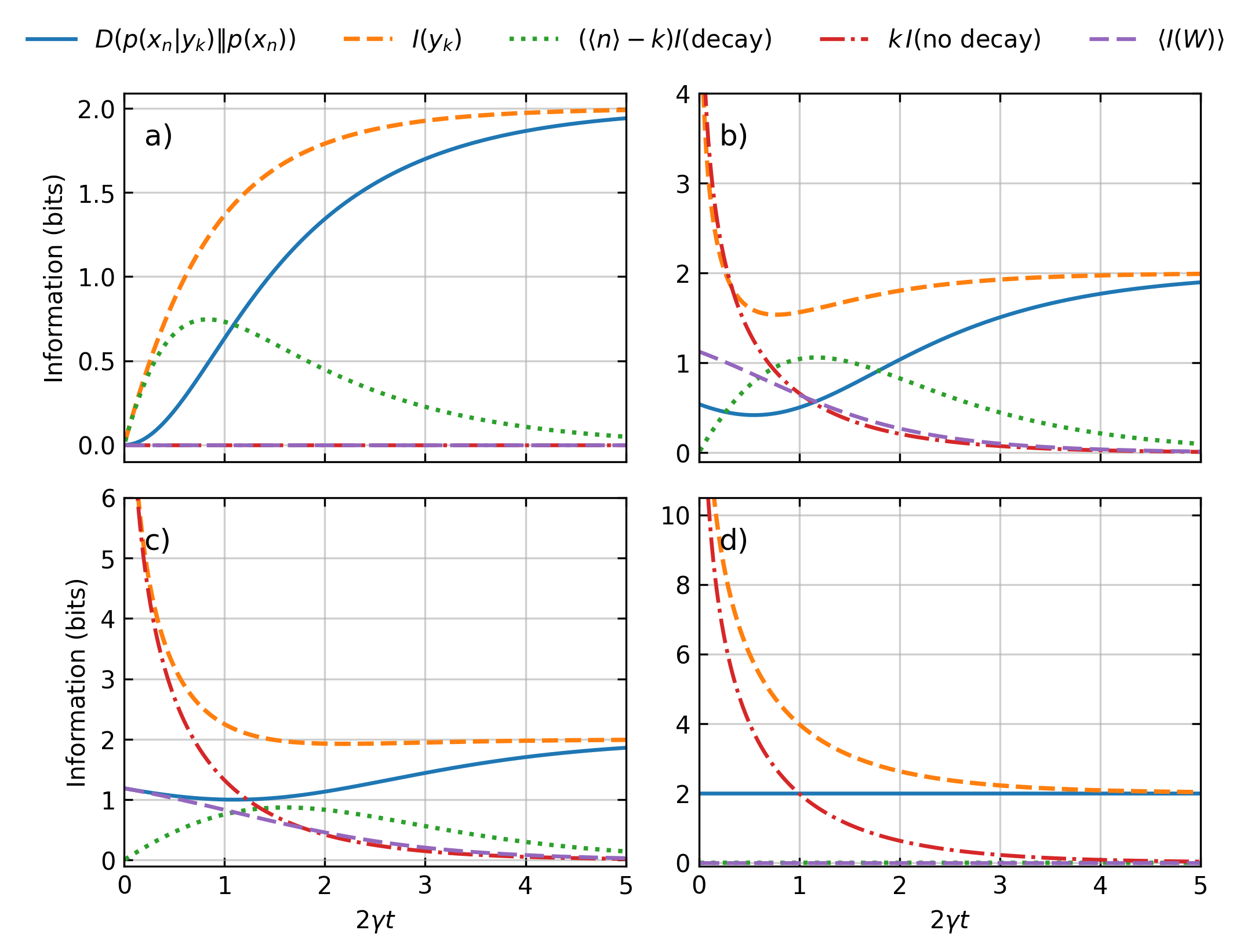}
  \caption{Informatics elements in Eq. \eqref{k-click}
  as functions of the rescaled time $2\gamma t$ for the state \(\ket{\psi}=c_0\ket{0}+c_1\ket{1}+c_2\ket{2}+c_3\ket{3}\). Each panel corresponds to the uniform prior distribution    
  $p=[1/4,\,1/4,\,1/4,\,1/4]$. The four panels correspond to detection outcomes 
  (a) $k=0$, (b) $k=1$, (c) $k=2$, and (d) $k=3$.}
  \label{fig:eq38_gain_loss}
\end{figure}

Hence, we arrive at a compact formula that encapsulates the information balance during a weak measurement with \(k\)-clicks. To illustrate this structure, Fig. \ref{fig:eq38_gain_loss} shows the four branches of information
as functions of \(2\gamma t\), for a four-level system with a uniform initial probability distribution, plotted for the cases \(k=0,1,2,3\).

In all panels, for large $\gamma t$, the information from the $k$-click outcome and the relative entropy are bounded by the prior’s total information (2 bits), in agreement with the conservation-type relations. For the null-result case ($k=0$, panel~(a)), these two quantities increase monotonically from zero, reflecting the initially maximal probability of a null detection, and then saturate. This behavior arises because the decay process drives the system toward the ground state. By contrast, the information associated with the decay process displays a transient peak and then vanishes as the posterior distribution concentrates on $\ket{0}$, as discussed earlier.

For single- and two-click outcomes [panels (b) and (c)], both the information from the $k$-click outcome in the detector and the no-decay process start at very large values. This is because the probability of a detection event is initially minimal, making the no-decay contribution dominant at short times. As time evolves, both quantities decrease monotonically: the no-decay contribution eventually vanishes, while the information gain first dips and then approaches the same saturation value as in the null-result case (2 bits). By contrast, the information associated with the decay process exhibits a peak, similar to the null-result case, and then decays to zero along with the no-decay term. The contribution from the distinguishability factor, $\langle I(W)\rangle$ reflects that a single- or two-click detection event does not perfectly distinguish among the underlying photon-number states that could have produced them, thereby reducing the effective information content. In this sense, $\langle I(W)\rangle$ represents a correction that accounts for the limited distinguishability of the detection outcome.
The relative entropy $D\!\big(p(x_n|y_k)\,\Vert\,p(x_n)\big)$ grows monotonically as the posterior departs from the prior for the null result case, while it decreases first and then increases in the case $k=1,2$. 

For three-click outcomes ($k=3$, panel~(d)), the qualitative behavior is similar, but the relative entropy is constant. In this case, however, the distinguishability correction \mbox{$\langle I(W)\rangle\equiv 0$} on the support ($n=3$), since $W=\binom{3}{3}=1$; the outcome, three clicks, fully specifies the photon number within the considered support, so no additional loss from this term occurs.

These trends follow directly from the structure of Eq. \eqref{decaylaw}. At short times, $p(y_k)$ is small for $k \geq 1$, so both the total information $I(y_k)$ and the no-decay contribution are large. At long times, the decay-related term approaches zero and sets $I(y_k) \approx -\log_2 p_k$. For the uniform prior considered here, this asymptotic value equals 2 bits. A clear distinction arises between the null result and the click outcomes, such that in the null-result case ($k=0$), $I(y_k)$ starts from zero and increases monotonically to its saturation value, whereas for $k\geq 1$ click events the gain always begins at a very large value and then decreases toward the same asymptotic limit.

\section{Conclusion}

We examined the way information in weak-measurement protocols is distributed among the measurement record, the system and the decay channel, and analyzed how this distribution is related to the success probability of weak-measurement reversal. For multilevel, decaying systems under continuous monitoring with dark detection, we derived an exact conservation relation for information. An analogous balance holds when an explicit reversal is applied, with the reversal success probability entering the relation.
 We further generalized the framework to finite-count detection (arbitrary photon numbers), obtaining quantitative trade-offs that link information content to various entities involved in the system. 
These results provide a unified framework of information flow in monitored open quantum dynamics and sharpen principles for analyzing and controlling weak measurements and their reversals in realistic noisy settings.

The framework invites several extensions such as understanding the information distribution and information flow in   Markovian and non-Markovian baths and structured environments, continuous-variable and Gaussian measurements,  measurement-based feedback and adaptive strategies, etc.

%\begin{acknowledgments}

%\end{acknowledgments}

%\bibliographystyle{unsrt}
\bibliography{refs}

\appendix

%\section{Appendixes}

%\nocite{*}


\end{document}